\documentclass[showpacs,10pt,twocolumn,prb]{revtex4-1}
\usepackage{amsmath}
\usepackage{amssymb}
\usepackage{graphics}
\usepackage{epsfig}
\usepackage{CJK}
\usepackage{color}

\begin{document}


\title{Type-I superconductivity in KBi$_{2}$ single crystals}
\author{Shanshan Sun$^{1,2}$, Kai Liu$^{1,2}$, and Hechang Lei$^{1,2,}$}
\email{hlei@ruc.edu.cn}
\affiliation{$^{1}$Department of Physics, Renmin University of China, Beijing 100872, China\\
$^{2}$Beijing Key Laboratory of Opto-electronic Functional Materials and Micro-nano Devices, Renmin University of China, Beijing 100872, China}
\date{\today}

\begin{abstract}

We report the detailed transport, magnetic, thermodynamic properties and theoretical calculation of KBi$_{2}$ single crystals in superconducting and normal states. KBi$_{2}$ shows metallic behavior at normal state and enters superconducting state below $T_{c}$ = 3.573 K. Moreover, KBi$_{2}$ exhibits low critical fields in all of measurements, field-induced crossover from second to first-order phase transition in specific heat measurement, typical magnetization isotherms of type-I superconductors, and small Ginzburg-Landau parameter $\kappa_{GL}=$ 0.611. These results clearly indicate that KBi$_{2}$ is a type-I superconductor with thermodynamic critical field $H_{c}=$ 234.3(3) Oe.

\end{abstract}

\maketitle

\section{Introduction}

Superconductivity is one of most attractive topics in condensed matter physics area not only because of the great application potential of those superconductors (SCs) with high transition temperature and critical field but also due to the importance of understanding the mechanism of Cooper pairing, especially for the unconventional superconductivity beyond Bardeen-Cooper-Schrieffer (BCS) mechanism. According to the Ginzburg-Landau (GL) theory, the value of GL parameter $\kappa_{GL}$ which is the ratio of penetration depth to coherence length classifies SCs into two categories: type-I SCs when $\kappa_{GL}<1/\sqrt{2}$ and type-II SCs when $\kappa_{GL}>1/\sqrt{2}$.\cite{Abrikosov} For most of superconducting compounds, they belong to the type-II SCs and have been studied extensively. In contrast, type-I SCs are thought empirically to occur mainly in elementary metals and metalloids and type-I superconducting compounds are very rare.\cite{Roberts} Recently, however, serval binary and ternary compounds are found to be type-I SCs, for instance, YbSb$_{2}$,\cite{Zhao,Yamaguchi} TaSi$_{2}$,\cite{Gottlieb} LaPd$_{2}$Ge$_{2}$,\cite{Hull} LaRh$_{2}$Si$_{2}$,\cite{Palstra} (Lu, Y, La)Pd$_{2}$Si$_{2}$,\cite{Palstra} LaRhSi$_{3}$,\cite{Anand} Ag$_{5}$Pb$_{2}$O$_{6}$,\cite{Yonezawa} ScGa$_{3}$ and LuGa$_{3}$.\cite{Svanidze} These studies break the empirical relation between type-I superconductivity and elemental metals and enlarge the family of type-I SCs to binary and ternary compounds.

For the binary bismuth compounds, some of them show superconductivity, such as KBi$_{2}$(3.6 K),\cite{Reynolds} SrBi$_{3}$ (5.62 K),\cite{Matthias} BaBi$_{3}$ (5.69 K),\cite{Matthias} Rh$_{3}$Bi$_{14}$ (2.94 K),\cite{ZhangX} and In$_{2}$Bi (5.9 K).\cite{Nishimura} For KBi$_{2}$, except superconducting transition temperature $T_{c}$, the studies on its physical properties are scarce and its classification of superconductivity has not been identified yet.\cite{Reynolds,Ponou} In this work, we performed the detailed characterization and analysis of physical properties for KBi$_{2}$ single crystals in superconducting and normal states. Experimental and theoretical calculation results undoubtedly indicate that KBi$_{2}$ is a type-I SC in the dirty limit. As far as we know, this is the first type-I SC in the bismuth compounds.

\section{Experimental}

Single crystals of KBi$_{2}$ were grown by the flux method with K : Bi = 1 : 9 molar ratio. K pieces (99.9 $\%$) and Bi shot (99.9 $\%$) were mixed and put into an alumina crucible, covered with quartz wool and then sealed into the quartz tube with partial pressure of Argon. The quartz tube was heated to 580 $^{\circ}$C for 12 h and then slowly cooled to 280 $^{\circ}$C where crystals were decanted with a centrifuge. Single crystals with typical size 1.8$\times$$1.8\times$1 mm$^{3}$ were obtained and exhibit metallic luster. X-ray diffraction (XRD) of powdered small crystals and a single crystal were performed using a Bruker D8 X-ray machine with Cu $K_{\alpha}$ radiation ($\lambda=$ 0.15418 nm) at room temperature. Rietveld refinement of the XRD patterns was performed using the code TOPAS4.\cite{TOPAS} Electrical transport, magnetization and specific heat measurements were performed in a Quantum Design PPMS-14. First-principles electronic structure calculations were carried out with the projector augmented wave method \cite{paw} as implemented in the VASP package.\cite{vasp} The generalized gradient approximation of Perdew-Burke-Ernzerh \cite{pbe} was adopted for the exchange-correlation potential. The kinetic energy cutoff of the plane-wave basis was set to be 350 eV. A supercell containing 2 K atoms and 4 Bi atoms and a $6\times6\times6$ k-point mesh for the Brillouin zone sampling were employed. The Gaussian smearing with a width of 0.05 eV was used around the Fermi surface. In structure optimization, both cell parameters and internal atomic positions were allowed to relax until the forces were smaller than 0.01 eV/\AA. The spin-orbital coupling (SOC) effect was included for the density of states (DOS) calculations.

\section{Results and discussion}

\begin{figure}[tbp]
\centerline{\includegraphics[scale=0.18]{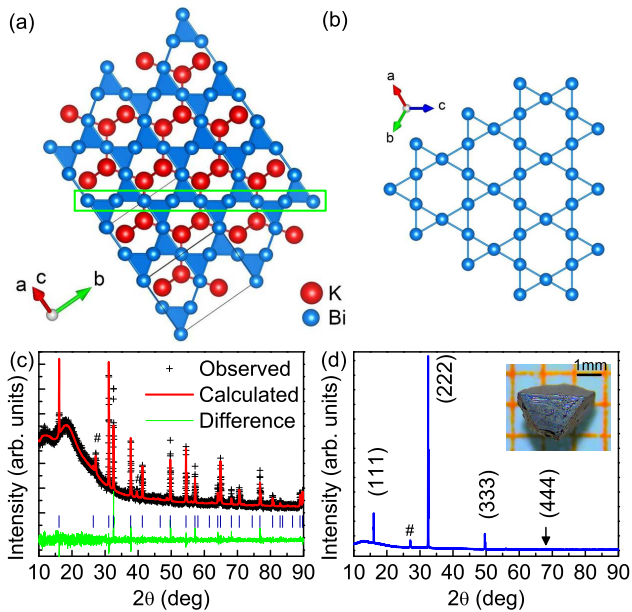}} \vspace*{-0.3cm}
\caption{(a) Crystal structure of KBi$_{2}$. The large red and small blue balls represent K and Bi atoms, respectively. The green rectangle marks the (111) plane. (b) The Kagom$\acute{e}$ net of Bi atoms in (111) plane. (c) Powder XRD pattern of KBi$_{2}$. (d) XRD pattern of a KBi$_{2}$ single crystal. The extra peaks in XRD patterns (labeled as the \# symbol) originate from residual Bi flux. Inset: photo of a typical KBi$_{2}$ single crystal. The length of one grid in the photo is 1 mm.}
\end{figure}

As shown in Fig. 1(a), KBi$_{2}$ is isostructural to MgCu$_{2}$ (Laves phase).\cite{Emmerling,Kuznetsov,Ponou} Each unit cell of KBi$_{2}$ has 8 K atoms and 16 Bi atoms. Every four Bi atoms form a tetrahedron and these tetrahedra connect each other by vertex-sharing to form a three-dimensional network. K atoms arrange in a diamond lattice which is intertwined with the network of Bi tetrahedra. On the other hand, it can be seen that there are two-dimensional (2D) Kagom\'{e} nets of Bi atoms in the (111) plane (Fig. 1(b)), which connect together along [111] direction by other Bi layers, and K atoms locate right above and below the center of hexagons of Bi atoms in the 2D Kagom\'{e} nets. The powder XRD pattern of KBi$_{2}$ can be well indexed using the $Fd\bar{3}m$ space group (MgCu$_{2}$-type structure) (Fig. 1(c)). The refined lattice parameters are $a=$ 0.95233(2) nm with $R_{p}=$ 5.88, $R_{wp}=$ 8.69 and $\chi^{2}=$ 1.28, which are close to the values reported in literature ($a=$ 0.95223(2)nm).$\cite{Ponou}$ There are some extra peaks that can be ascribed to the diffraction of residual Bi flux. The XRD pattern of a KBi$_{2}$ single crystal indicates that the surface of crystal is parallel to the (111) plane (Fig. 1(d)). The KBi$_{2}$ crystals prefer to form triangle surface as shown in the inset of Fig.1(d). This characteristic is consistent with the single crystal XRD pattern as well as its crystallographic symmetry.

\begin{figure}[tbp]
\centerline{\includegraphics[scale=0.23]{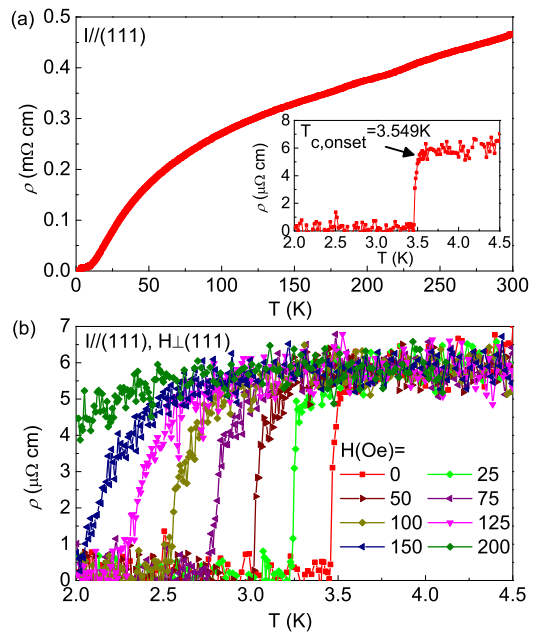}} \vspace*{-0.3cm}
\caption{(a) Temperature dependence of electrical resistivity $\rho(T)$ of KBi$_{2}$ single crystal at zero field. Inset: expanded view of the low-temperature data showing the superconducting transition. The $T_{c,onset}$ at zero field is marked by arrow. (b) Low-temperature dependence of $\rho(T)$ of KBi$_{2}$  single crystal at various magnetic fields from 0 to 200Oe.  }
\end{figure}

Fig. 2(a) shows the temperature dependence of electrical resistivity $\rho(T)$ for KBi$_{2}$ single crystal at zero field. It can be seen that KBi$_{2}$ exhibits metallic behavior and the curve $\rho(T)$ is convex above 25 K, with a tendency to saturate at high temperature. This is a typical shape of resistivity for the sample in which the dominant scattering mechanism is electron-phonon scattering. The residual resistivity ratio (RRR), defined as $\rho$(300 K)/$\rho$(4 K), is about 72.5, indicating the high quality of single crystal. On the other hand, at zero field, there is a sharp superconducting transition in $\rho(T)$ curve with $T_{c,onset}=$ 3.549 K and transition width $\Delta T_{c}\sim$ 0.1 K (inset of Fig. 2(a)). The transition temperature is consistent with previous magnetic susceptibility measurement ($T_{c}=$ 3.5 K).\cite{Ponou} Moreover, with increasing magnetic field, the superconducting transition shifts to lower temperatures gradually and the transition width also becomes wider for $H\perp$ (111) (Fig. 2(b)). Surprisingly, at very low field ($H=$ 200 Oe), the $T_{c}$ has been suppressed below 2 K.

\begin{figure}[tbp]
\centerline{\includegraphics[scale=0.15]{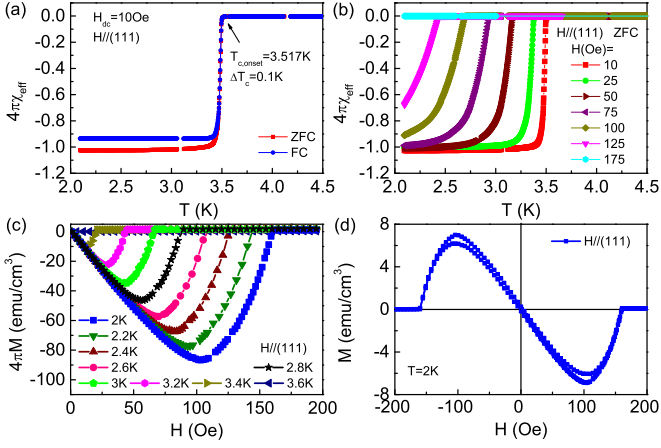}} \vspace*{-0.3cm}
\caption{(a) Temperature dependence of magnetic susceptibility $\chi_{eff}(T)$ = $\chi(T)$/(1-$N_{d}$$\chi$) scaled by 4$\pi$ at $H=$ 10 Oe with ZFC and FC modes. The $T_{c,onset}$ is marked by arrow. (b) Temperature dependence of $4\pi\chi_{eff}(T)$ at various magnetic fields with ZFC mode. (c) Low-field parts of $M(H)$ at various temperatures with demagnetization correction.(d) Magnetization loop at $T=$ 2 K. For all of measurements, the field direction is parallel to the (111) plane of KBi$_{2}$ single crystal.}
\end{figure}

As shown in Fig. 3(a), low-temperature magnetic susceptibility of KBi$_{2}$  single crystal for $H\parallel$ (111) with $H=$ 10 Oe exhibits diamagnetic signals with sharp transitions appearing at $T_{c,onset}=$ 3.517 K ($\Delta T_{c}=$ 0.10 K) for both the zero field cooling (ZFC) and field cooling (FC) modes. This superconducting transition temperature is almost same as that obtained from transport measurement, confirming the superconducting ground state of KBi$_{2}$. After considering the demagnetization effect of sample by using the formula $4\pi\chi_{eff}=4\pi\chi/(1-N_{d}\chi)$ where $N_{d}$ is demagnetization factor and the calculated value is about 0.222,\cite{Aharoni} the superconducting volume fraction estimated from the ZFC data is very close to 100 \%, indicating that the bulk superconductivity in KBi$_{2}$. Meanwhile, the large superconducting volume fraction for FC data suggests the flux pinning effect is rather weak or even absent. Further increasing field suppresses the $T_{c}$ gradually and when $H=$ 175 Oe, the normal states behavior persists down to 2 K, similar to the trend in $\rho(T)$ curves (Fig. 2(b)).

Fig. 3(c) shows the dc magnetization isotherms $4\pi M(H)$ of KBi$_{2}$ single crystal at various temperatures (from 2 K to 3.6 K) for $H\parallel $ (111) at the low-field range. It can be seen that the magnetic fields corresponding to the maximum absolute values of magnetic moment are very close to those where $4\pi M(H)$ become zero. Based on this, two different assumptions can be made: either (i) KBi$_{2}$ is a type-II SC that the upper critical field $H_{c2}$ is very small and close to the lower critical field $H_{c1}$; or (ii) KBi$_{2}$ is a type-I SC and the departure from the ideal step-like transition at critical field may be attributed to a strong pinning of domain walls in the intermediate state or to unclear demagnetization.$\cite{Tran}$ For the latter case, the critical field $H_{c}$ is defined as the field where the sample enters to the normal state ($M(H)=$ 0). On the other hand, when temperature decreases, the critical field shifts to higher field. Even at the lowest measuring temperature (2 K), however, the critical field is still remarkably small ($\sim$ 160 Oe). The full magnetization loop of KBi$_{2}$ measured at 2 K for $H\parallel$ (111) is shown in Fig. 3(d). It can be seen that the hysteresis is very small. The small critical field accompanying with the weak magnetization hysteresis and the shape of $M(H)$ loop similar to other type-I superconducting compounds, such as YbSb$_{2}$,\cite{Zhao,Yamaguchi} LaRhSi$_{3}$,\cite{Anand}  ScGa$_{3}$ and LuGa$_{3}$,\cite{Svanidze} suggest that the KBi$_{2}$ is more like a type-I SC rather than a type-II SC.

\begin{figure}[tbp]
\centerline{\includegraphics[scale=0.23]{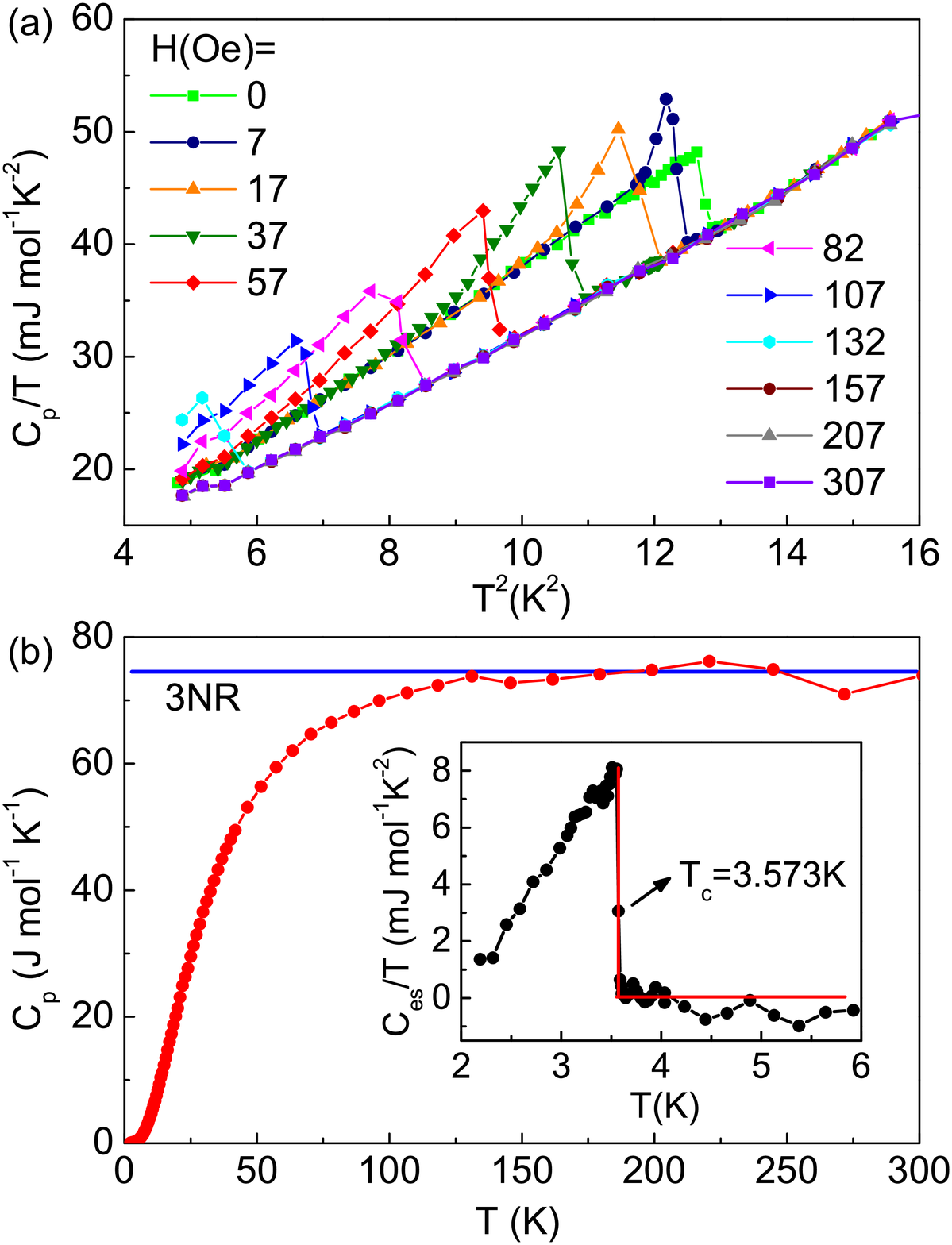}} \vspace*{-0.3cm}
\caption{(a) Low-temperature specific heat $C_{p}/T$ vs. $T^{2}$ of KBi$_{2}$ single crystal in a series of magnetic fields from 0 to 307 Oe. (b) Temperature dependence of specific heat from 2.2 K to 300 K at zero field. The solid line shows the classic value of specific heat at high temperature limit. Inset: Temperature dependence of the electronic specific heat plotted as $C_{es}/T$ vs. $T$ at zero field.}
\end{figure}

Specific heats of KBi$_{2}$ single crystal at various fields are shown in Fig. 4(a). At zero field, there is a jump at 3.573 K, indicating the bulk superconductivity of KBi$_{2}$. The transition temperature is also consistent with the $T_{c}$ obtained from resistivity and magnetization measurements. More importantly, the specific-heat jump at superconducting transition at $H=$ 7 Oe is much sharper and higher than that at zero field, which is a unique feature of type-I SCs and indicates that there is a crossover from second- to first-order phase transition when field is applied. This behavior has been observed in other type-I intermetallic SCs, such as YbSb$_{2}$,\cite{Zhao,Yamaguchi} LaRhSi$_{3}$,\cite{Anand}  ScGa$_{3}$ and LuGa$_{3}$,\cite{Svanidze} Therefore, specific heat measurement further supports that the KBi$_{2}$ is a type-I SC. On the other hand, the superconducting transition shifts to lower temperature with increasing field and the superconductivity is suppressed below 2.2 K when $H>$ 157 Oe. In the normal state ($H =$ 307 Oe ), the electronic specific heat coefficient $\gamma$ and phonon specific heat coefficient $\beta$ are obtained using the linear fit $C_{p}/T=\gamma+\beta T^{2}$. The fitted $\gamma$ and $\beta$ is 1.3(4) mJ mol$^{-1}$ K$^{-2}$ and 3.10(4) mJ mol$^{-1}$ K$^{-4}$, respectively. The latter one gives the Debye temperature $\Theta _{D}=$ 123.4(5) K using the formula $\Theta_{D}=(12\pi ^{4}NR/5\beta )^{1/3}$. The electron-phonon coupling $\lambda _{e-ph}$ can be estimated with the values of $\Theta _{D}$ and $T_{c}$ using McMillan's theory,\cite{McMillan}

\begin{equation}
\lambda _{e-ph}=\frac{1.04+\mu ^{\ast }\ln(\Theta _{D}/1.45T_{c})}{(1-0.62\mu ^{\ast })\ln(\Theta _{D}/1.45T_{c})-1.04}
\end{equation}

where $\mu^{\ast}$ is the repulsive screened Coulomb potential and is usually between 0.1 and 0.15. Setting $\mu ^{\ast}=$ 0.13, the calculated $\lambda _{e-ph}$ is 0.774, implying that KBi$_{2}$ is an intermediately or strongly coupled BCS SC. Fig. 4(b) shows the specific heat of KBi$_{2}$ measured from 2.2 K to 300 K at zero field. The specific heat at high temperature approaches the value of $3NR$ at 300 K, where $N$ is the atomic number in the chemical formula ($N=$ 3) and $R$ is the gas constant ($R=$ 8.314 J mol$^{-1}$ K$^{-1}$), consistent with the Dulong-Petit law. The electronic specific heat $C_{es}$ at zero field (inset of Fig. 4(b)) is obtained by subtracting the lattice contribution from the total specific heat. The extracted electronic specific heat jump at $T_{c}$ ($\Delta C_{es}/\gamma T_{c}=$ 6.06) is much larger than the weakly coupled BCS value 1.43, indicating the strongly coupled superconductivity in KBi$_{2}$.\cite{McMillan}

The thermodynamic critical field $H_{c}(T)$ can be obtained by integrating the differences of specific heats and the specific heats divided by temperature between the zero-field superconducting (s) and normal (n) states ($H=$ 207 Oe) (free energy analysis),\cite{Anand,Wang}

\begin{equation}
- H_{c}(T)^{2}/8\pi = \Delta F(T) = \Delta U(T) - T\Delta S(T)
\end{equation}
\begin{equation}
\Delta U(T) = \int_T^{T_{c}} [C_{s}(T') - C_{n}(T')] dT'
\end{equation}
\begin{equation}
\Delta S(T) = \int_T^{T_{c}} \frac{C_{s}(T') - C_{n}(T')}{T'} dT'
\end{equation}

where $\Delta F(T)$, $\Delta U(T)$, and $\Delta S(T)$ are the differences of free energy, internal energy, and entropy between zero-field superconducting and normal states. The relationship between the $T_{c}$ and critical fields determined from the magnetization, resistivity, specific heat measurements and free energy analysis is summarized in the $H-T$ phase diagram (Fig. 5). It can be seen that all of data are almost on one line, confirming the consistency of critical fields from different measuring methods and undoubtedly indicating that all of these critical fields are thermodynamic critical field $H_{c}(T)$. The dot line gives the free energy fitting that using the BCS temperature dependence $H_{c}(T)=H_{c}(0)[1-(T/T_{c})^{2}]$. This gives the zero-temperature $H_{c}(0)=$ 234.3(3) Oe and $T_{c}$ = 3.561(1) K, close to those values obtained from different experimental methods.

\begin{figure}[tbp]
\centerline{\includegraphics[scale=0.28]{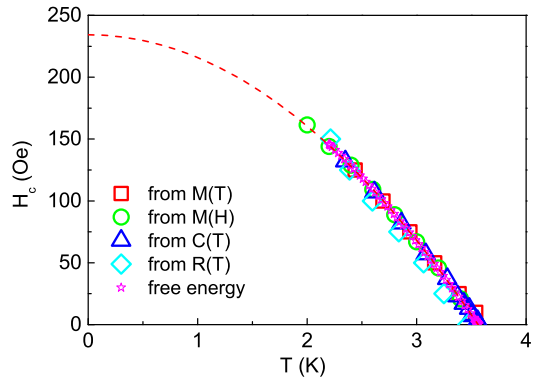}} \vspace*{-0.3cm}
\caption{$H-T$ phase diagram for KBi$_{2}$ determined from $M(T)$ (squares), $R(T)$ (triangles), and $C_{p}(T)$ (rhombuses) at various fields, $M(H)$ at various temperatures (circles) as well as that calculated from free energy analysis from zero-field specific heat data (stars). The dot line is fitted by using $H_{c}(T)=H_{c}(0)[1-(T/T_{c})^{2}]$.}.
\end{figure}

Fig. 6 shows the calculated density of state (DOS) of KBi$_{2}$. The shape of DOS is somewhat different from previous result,\cite{Ponou} because the SOC which is significant for bismuth is considered in present calculation. The finite DOS at Fermi energy level ($E_{F}$) indicates the metallic ground state, consistent with the experimental results. The DOS near $E_{F}$ is mainly contributed by Bi-$p$ state and the contribution of K is negligible, indicating that K atoms transfer almost all of valence electron to the Bi atoms, i.e. the valence state of K is +1 and the three-dimensional network of Bi can be thought as polyanion.\cite{Ponou}

\begin{figure}[tbp]
\centerline{\includegraphics[scale=0.29]{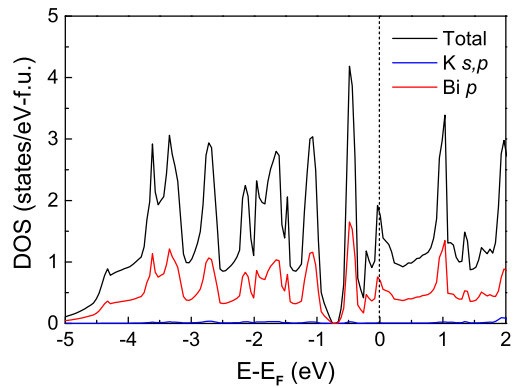}} \vspace*{-0.3cm}
\caption{Calculated density of states (DOS) of KBi$_{2}$. Black, red and blue lines correspond to the total DOS, partial DOS of Bi-$p$ state and partial DOS of K-$s,p$ state, respectively. The dotted line represents the Fermi energy level, $E_{F}$.}
\end{figure}

\begin{table}[tbp]\centering
\caption{Physical parameters of KBi$_{2}$ in superconducting and normal states.}
\begin{tabular}{ccccccccccccccccc}
\hline\hline

 Parameter                    &&&&&&&&&&&&&&&& Value\\\hline
 $T_{c}$ (K)                  &&&&&&&&&&&&&&&& 3.573\\
 $H_{c}$ (Oe)                 &&&&&&&&&&&&&&&& 234.3(3)\\
 RRR                          &&&&&&&&&&&&&&&& 72.5\\
 $\gamma$ (mJ/mol K$^{2}$)    &&&&&&&&&&&&&&&& 1.3(4)\\
 $\Theta _{D}$ (K)            &&&&&&&&&&&&&&&&  123.4(5)\\
 $\lambda _{e-ph}$            &&&&&&&&&&&&&&&&  0.774  \\
 $\Delta C_{es}/\gamma T_{c}$ &&&&&&&&&&&&&&&& 6.06 \\
 m* ($m_{e}$)                 &&&&&&&&&&&&&&&& 0.312\\
 $\lambda_{L}(0)$ (nm)        &&&&&&&&&&&&&&&& 11.8\\
 $l_{tr}$ (nm)                &&&&&&&&&&&&&&&& 13.9 \\
 $\xi_{GL}(0)$ (nm)           &&&&&&&&&&&&&&&& 133.7\\
 $\lambda_{GL}(0)$ (nm)       &&&&&&&&&&&&&&&&  81.7\\
 $\kappa_{GL} $               &&&&&&&&&&&&&&&& 0.611\\
\hline\hline
\end{tabular}
\label{TableKey copy(1)}
\end{table}

The calculated total DOS at $E_{F}$ is 1.77 states/eV-f.u. and it gives the carrier density $n$ = 6.33$\times$10$^{22}$ cm$^{-3}$ using the free-electron model. The calculated Fermi wave vector $k_{F}$ equals 12.33 nm$^{-1}$ using the formula $k_{F}=(3\pi^{2}n)^{1/3}$. Then, the effective electron mass can be determined as $m^{*}=3\hbar^{2}\gamma/(k_{B}^{2}k_{F}V)=$ 0.312 $m_{e}$, where $k_{B}$ is Boltzmann constant and $m_{e}$ is the free electron mass. With derived $m^{*}$ and $n$, the London penetration depth is calculated as $\lambda_{L}(0)=(m^{*}/\mu_{0}ne^{2})^{1/2}=$ 11.8 nm.\cite{Orlando} Meanwhile the coherence length is determined by using the BCS relation $\xi(0)=0.18\hbar^{2}k_{F}/(k_{B}T_{c}m^{*})$,\cite{Orlando} which gives $\xi(0)=$ 1.76 $\mu$m. Assuming a simple model of a spherical Fermi surface ($S/S_{F}=$ 1), the mean free path $l_{tr}$ is estimated as $l_{tr}= 1.27\times10^{8}(\rho_{0}n^{2/3}S/S_{F})^{-1}=$ 13.9 nm, where $\rho_{0}$ is the low-temperature normal state resistivity (5.76 $\mu\Omega$ cm at 4 K).\cite{Orlando} It clearly indicates that the electronic mean free path is considerably smaller than the BCS coherence length ($l_{tr}/\xi(0)=$ 0.008), suggesting that KBi$_{2}$ can be classified as a SC in the dirty limit. In the dirty limit, the GL parameter $\kappa_{GL}= 0.72\lambda_{L}(0)/l_{tr}=$ 0.611,\cite{Orlando} which is smaller than $1/\sqrt{2}$, further confirming that KBi$_{2}$ is a type-I SC. Moreover, the zero-temperature Ginzburg-Landau (GL) coherence length in the dirty limit $\xi_{GL}(0)$ can be obtained from the relations $\xi_{GL}(0)= 8.57\times10^{-9}(10\gamma\rho_{0}T_{c}/V)^{-1/2}$ .\cite{Orlando} It gives $\xi_{GL}(0)=$ 133.7 nm. According to the definition of $\kappa_{GL}=\lambda_{GL}(0)/\xi_{GL}(0)$, the derived zero-temperature GL penetration depth $\lambda_{GL}(0)$ is 81.7 nm. The superconducting and thermodynamic parameters of KBi$_{2}$ are summarized in Table I.

\section{Conclusion}

In summary, the single crystal of KBi$_{2}$ has been grown from Bi flux successfully. Resistivity, magnetization and specific heat measurements indicate that KBi$_{2}$ shows a bulk superconductivity with $T_{c}$ = 3.573 K. Further analysis of experimental results indicate that KBi$_{2}$ is type-I BCS SC in the dirty limit with an intermediate or strong coupling strength. The thermodynamic critical field is 234.3(3) Oe and the calculated GL parameter $\kappa_{GL}=$ 0.611. This study not only deepens our understanding on type-I superconductivity but will also stimulate further work on discovering other type-I SCs in binary or ternary compounds.

\section{Acknowledgments}

This work was supported by the Ministry of Science and Technology of China (973 Project: 2012CB921701), the Fundamental Research Funds for the Central Universities, and the Research Funds of Renmin University of China (RUC) (15XNLF06 and 14XNLQ03), and the National Natural Science Foundation of China (Grant No. 11574394). Computational resources have been provided by the PLHPC at RUC.

\end{document}